# Scaling Correlations Among Central Massive Objects and Their Host Galaxies

I. Tosta e Melo[1], R. Capuzzo-Dolcetta[1],

[1]Sapienza-Università di Roma, P.le A. Moro 5, I-00185 Roma, Italy

E-mail: `iara.tosta.melo@gmail.com`

**Abstract.** The central regions of galaxies show the presence of super massive black holes and/or very dense stellar clusters. Both such objects seem to follow similar host-galaxy correlations, suggesting that they are members of the same family of Compact Massive Objects. Here we investigate a huge data collection of Compact Massive Objects properties to correlate them with absolute magnitude, velocity dispersion and mass of their host galaxies. We draw also some preliminary astrophysical conclusions.

## 1. Introduction

The link between the formation and evolution of galaxies and those of their central region is a debated topic. Various studies suggest that massive galaxies, both elliptical and spiral, harbor a SuperMassive Black Hole (SMBH) in their centers, with masses between $10^6 - 10^9$ M$_\odot$. The SMBH masses correlate with various properties of their host galaxies, such as the bulge luminosity [1], mass [2], velocity dispersion [3], and light profile concentration [4].

Galaxies across the entire Hubble sequence also show the presence of massive and compact stellar clusters referred to Nuclear Star Clusters (NSCs). NSCs are bright (4 mag brighter than an ordinary globular cluster), massive ($10^6 - 10^7$ M$_\odot$) and very dense systems, with half light radius of 2-5 pc [5]. In elliptical galaxies, the NSCs are also referred to as Resolved Stellar Nuclei (RSN). All the NSCs contain an old stellar population (age > 1 Gyr), and, in some cases, show the presence of a young stellar population (age < 100 Myr) [6].

The models of formation and evolution of NSCs are still under debate.

Ref. [7] showed that, despite their different morphologies, some galaxies of the Local Group present both a NSC and a SMBH. In such case, the SMBH is surrounded by the NSC. Those objects follow a similar host-galaxy correlation [8], suggesting that they are members of the same family of Compact Massive Objects (CMOs). CMOs constitute a interplay between either SMBH or a compact stellar structure (NSCs, Nuclear Stellar Disk or Resolved Stellar Nuclei).

Ref. [9] found a separation in mass between galaxies that host a SMBH (those with $M > 5 \times 10^9$ M$_\odot$) and galaxies that host a NSC (with $M < 5 \times 10^9$ M$_\odot$). A transition region exists between $10^8$ and $10^{10}$ M$_\odot$ in which NSCs and SMBHs can coexist [10, 11].

A good example is the Milky Way (MW), where a $4 \times 10^6$ M$_\odot$ black hole coexists with an NSC~4 times more massive ($M_{NSC} \approx 1.5 \times 10^7$ M$_\odot$).

Studies in the literature also showed that the NSC mass versus the host galaxy velocity dispersion ($\sigma$) relation is roughly the same observed for SMBHs. Ref. [12] claimed, instead, that the $M_{NSC} - \sigma$ relation is shallower for NSCs ($M_{NSC} \propto \sigma^{1.5}$) than for SMBHs. Moreover, it has been shown that the NSC masses ($M_{NSC} \propto \sigma^{1.5}$) correlate better with the bulge mass, while the SMBH mass displays a tighter correlation with the total mass of the galaxy [13].

Here we investigate these scaling correlations for a set of NSC and SMBH data wider than what already studied in the literature.

The paper is organized as follows: Section 2 presents the data base used for building our sample, and Section 3 describes the methodology. The results and their discussion are presented in Section 4.

## 2. The database

Ref. [14] selected 43 galaxies of the Fornax Cluster with early-type morphologies (E, S0, SB0, dE, dE,N or dS0,N), using the $F475W$ and $F850LP$ bandpasses of the Hubble Space Telescope (HST) Advanced Camera of Survey (ACS). 31 galaxies out of 43, representing $72 \pm 13\%$ of their sample, are nucleated with the majority of the nuclei bluer than their host galaxies. This approach gives two apparent magnitudes for the nuclei, in the $g$ and $z$ band. The $g$ band values were in this work, because the level of nucleation is slightly larger in this band.

Ref. [15] analysed the nuclei of a sample of 100 early-type galaxies in the Virgo cluster of morphological types E, S0, dE, dEN and dS0. The images were taken with the ACS instrument in Wide Field Channel (WFC) using a combination of the F475W and F850LP filters, roughly equivalent to the $g$ and $z$ bands, in the Sloan Sky Survey photometric system. The authors also concluded that nucleated galaxies are more concentrated toward the center of Virgo cluster. Some nuclei of ACS Virgo Cluster Survey (ACSVCS) are bluer than the underlying galaxies and a central excess is more apparent in the $g$ band rather than in the redder bandpass.

Ref. [16] presented the properties of 228 nuclear stellar cluster in nearby late-type disk galaxies observed with the WFPC2/Hubble Space Telescope (HST/WFPC2), in $B$ and $I$ bands, with distance $\leq 40$ Mpc and distance modulus $\leq 33$ mag. The authors avoided the most luminous bulges and all ANGs to build the sample, because the presence of strong ANGs would complicate the NSC characterization. They also concluded that largest and brightest NSCs occupy the regime between Ultra Compact Dwarf (UCD) and early-type galaxies.

Ref. [17] analysed the light profile of 200 early-type dwarf galaxies with magnitudes $16.0 < m_{F814W} < 22.6$ mag using the HST/ACS Coma Cluster Survey. NSCs are detected in 80% of the galaxies, thus doubling the sample of HST-observed early-type dwarf galaxies with nuclear star clusters.

We also included in our sample 89 galaxies having a dynamical detection of the central black hole mass. We took information from a catalog presented by [18], and also from a sample presented by [19].

## 3. Method

Our aim was to estimate CMOs masses for each of the catalogues presented in the previous section of the paper, and compare such values with the mass, velocity dispersion and absolute B magnitude of their host galaxies.

The stellar M/L ratio vs color correlation given by [20] was used to determine the CMOs masses.

The CMOs mass values of ACS Fornax and Virgo Cluster Surveys were obtained by the $g - z$ color correlation. Georgiev's work did not provide the $B - V$ color correlation (reddening corrected), for the HST/WFPC2 archive nuclei. Such values were provided in the Johnson/Cousin photometric system and our reference system is the AB one. We used the correlation given by [21] to do the conversion:

$$B = B(AB) + 0.163(\pm 0.004), \qquad (1)$$

$$V = V(AB) + 0.044(\pm 0.004). \qquad (2)$$

The magnitudes for the NSCs given for Coma Cluster are only in the $F814W$ band, which is equilavent to the $I_C$ band ($AB$ photometry system). The authors did not provide the color-correlation for this sample. We used the CMOs color-correlation average of the previous sample to get our final results.

The galaxy masses for the Fornax and Virgo clusters were computed using the virial theorem:

$$M_{\text{gal}} = \frac{\beta R_{\text{eff}} \sigma^2}{G}, \qquad (3)$$

where $G$ is the gravitational constant, $\sigma$ is the velocity dispersion, $R_{\text{eff}}$ is the effective radius of the galaxies, and $\beta = 5$.

For the ACS Fornax Cluster Survey the effective radius values for the galaxies used were provided by [22]. There are no available estimates of the effective radius for FCC 2006, FCC 1340, and FCC 21. The values of the effective radius for galaxies in the ACSVCS were given in their respective catalog.

To compute the galaxy masses in the HST/WFCP2 archive we used the same magnitude vs $B-V$ color correlation for the galaxies and, after the conversion between the reference magnitude systems, from Johnson/Cousin photometric system into the AB system for both bands (E.1 and E.2), we obtained our final results using the stellar M/L ratio vs color correlation formula.

The values for the effective radii are not provided in the ACS Coma Cluster (ACSCCS) catalog. Thus, for it we used the galaxies color-correlation average of the previous sample and we applied such values in stellar M/L ratio vs color correlation formula.

The velocity dispersion values for the galaxies in the Fornax and Virgo Cluster, and those in the HST/WFCP2 archive were taken from the Hyperleda Website[1]; for the Coma Cluster the values are given by [23, 24].

Ref. [14, 16] work also provide the values for the apparent B magnitude for the galaxies in Fornax Cluster and HST/WFCP2 archive, respectively. Such values, for the Virgo Cluster, were taken from [6]. For the Coma Cluster we converted the values of apparent $I_C$ magnitude into apparent $B$ magnitude:

$$F475W - F814W = -0.036(F475W + 18) + 1.13, \qquad (4)$$

where $F475W$ and $F814W$ are equivalent to B and $I_C$ bands, respectively.

All the values for the SMBH database were taken in the literature, such as absolute B magnitude [25, 26], velocity dispersion [27], and masses of their host galaxies [25].

## 4. Comparative discussion of results

Our results are plotted in Figs. 1-4. The existence of evident correlations indicates a direct link among large galactic spatial scales and the much smaller scales of nuclear environment. We have fitted scaling correlations connecting the CMOs mass to various properties of their host galaxies: $M_{\text{B}}$, $\sigma$, and galaxy mass. To obtain the final value of the two fit parameters, $a$ and $b$, for each galaxy cluster, we applied a linear best fit, i.e. least $\chi^2$ method. Thus the values shown in Table 1 represent the best approximation recoverable.

NSCs are more common in fainter galaxies, with magnitudes between $-21 \leq M_{\text{B}} \leq -13$, than in brighter galaxies as shown in Figs. 1 and 4. The same behaviour was already predicted by [15].

The presence of NSCs is much more frequent in galaxies with lower velocity dispersion, $\sigma \leq 100$ km/s (Fig. 2).

---

[1] http://leda.univ-lyon1.fr/

**Table 1.** Valeus of the best least square fits. In the first column numbers 1, 2, 3 and 4 represent the ACSFCS, ACSVCS, HST/WFCP2 and ACSCCS samples respectively. The slope $a$ and linear coefficient $b$, with their respectives errors for the logarithmic fit of $M_{CMO}$ as function of $M_B$, $\sigma$, and the mass of the host galaxies are reported in the other columns.

| Sample | $M_B$ (mag) | | $\sigma$ (km/s) | | $M_{\text{galaxies}}/M_{\text{sun}}$ | |
|---|---|---|---|---|---|---|
| | $a$ | $b$ | $a$ | $b$ | $a$ | $b$ |
| 1 | $-0.57 \pm 0.09$ | $-2.71 \pm 1.58$ | $1.94 \pm 0.81$ | $3.76 \pm 1.49$ | $0.95 \pm 0.27$ | $-2.07 \pm 2.7$ |
| 2 | $-0.47 \pm 0.04$ | $-1.02 \pm 0.81$ | $2.28 \pm 0.25$ | $3.35 \pm 0.49$ | $0.75 \pm 0.09$ | $0.12 \pm 0.92$ |
| 3 | $-0.28 \pm 0.04$ | $0.45 \pm 0.89$ | $0.54 \pm 0.56$ | $5.17 \pm 0.99$ | $0.37 \pm 0.11$ | $2.63 \pm 1.01$ |
| 4 | $-0.14 \pm 0.04$ | $3.88 \pm 0.66$ | $1.49 \pm 0.88$ | $1.64 \pm 1.40$ | $0.49 \pm 0.06$ | $2.43 \pm 0.45$ |

Galaxies with larger masses have more massive CMOs (Fig. 3). It is likely that MBHs and compact stellar structures in these galaxies interacted in the same way during their formation.

## 5. Conclusions

We studied three different scaling correlations with an updated set of data involving CMOs mass and host galaxies properties. NSCs (or CMOs and SMBH) appear to not to follow the same scaling relation respect to the host galaxy mass, in agreement with the previous studies. Our list of findings are:

(i) Galaxies brighter than $M_B = -18$ host SMBHs, and the existence of such objects in bright galaxies reconcile with the existence, in most of the cases, of an AGN. The lack of NSCs in faint galaxies may be related to the small number of globular clusters in some galaxies, as in the Local Group, where globular clusters are not found in stellar spheroids fainter than $M_B \sim -12.5$.

(ii) We reconfirmed that, with our updated set of data, as one moves to fainter galaxies, the nuclei become the dominant feature while MBHs tend to become less common and, perhaps, entirely disappear at the fainter end [10].

(iii) The $M_{NSCs}$–$\sigma$ correlation shows a slope between 1.5 and 3, which is in good agreement with that obtained by [1], excluding HST/WFCP2 archive sample.

(iv) It is unclear what exactly the result of any interaction may be: (1) the presence of a single SMBH may evaporate the compact stellar structure, (2) a single or binary SMBH may heat and erode the surrounding compact stellar structure.

**Acknowledgments**
I. Tosta e Melo acknowledges CAPES-Brazil for support through the grant 9467/13-0.

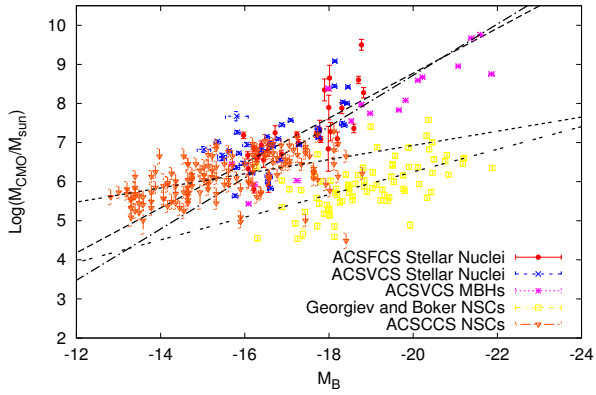

**Figure 1.** Masses of the CMOs versus the absolute B magnitude of the host galaxies. The black dashed line, the black dashed pointed line, the black two pointed line, and the black pointed line are the least square fit of the Fornax, Virgo, Georgiev and Boker, and Coma sample, respectively.

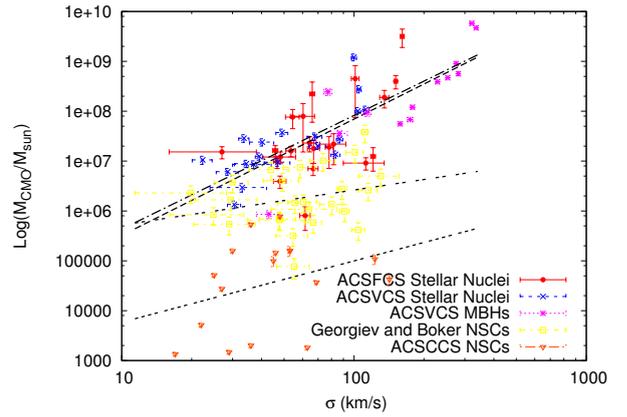

**Figure 2.** Masses of the CMOs versus the velocity dispersion of the host galaxies. The legenda for the various least square fits is the same of Fig. 1

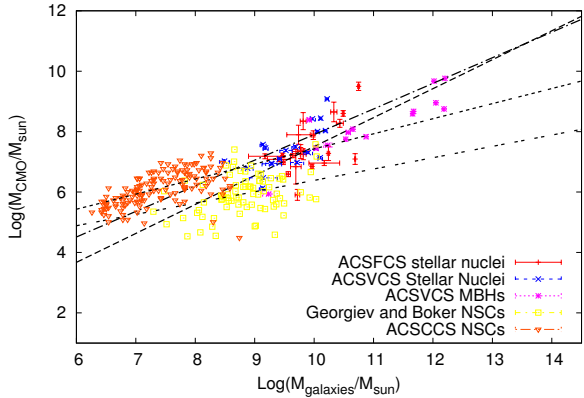

**Figure 3.** Masses of the CMOs versus the mass of the host galaxies. The legenda for the various least square fits is the same of Fig. 1

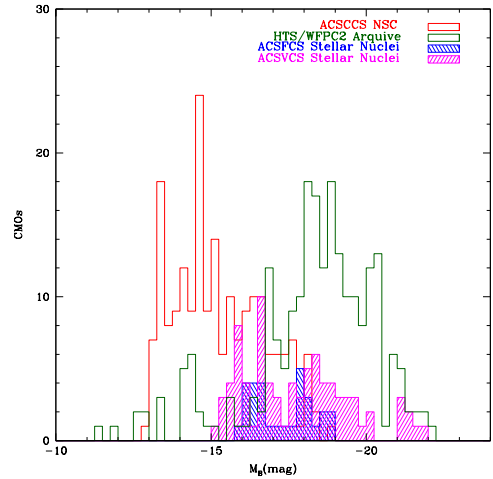

**Figure 4.** Distribution of the B integrated magnitudes of CMOs for the 571 objects belonging to the ACSFCS, ACSVCS, HST/WFCP2 archive and ACSCCS sample.